*Research Article*

# Global Distribution of Google Scholar Citations: A Size-independent Institution-based Analysis

Aparna Basu, Deepika Malhotra, Taniya Seth, Pranab Kumar Muhuri*

*Department of Computer Science, South Asian University, Akbar Bhawan, Chanakyapuri, New Delhi, INDIA.*

**ABSTRACT**
Most currently available schemes for performance-based ranking of Universities/Research organizations, such as, Quacarelli Symonds (QS), Times Higher Education (THE), Shanghai University-based All Research of World Universities (ARWU) use a variety of criteria that include productivity, citations, awards, reputation, etc., while Leiden and Scimago use only bibliometric indicators. The research performance evaluation in the aforesaid cases is based on bibliometric data from Web of Science or Scopus, which are commercially available priced databases. The coverage includes peer-reviewed journals and conference proceedings. Google Scholar (GS) on the other hand, provides a free and open alternative to obtaining citations of papers available on the net, (though it is not clear exactly which journals are covered.) Citations are collected automatically from the net and also added to self-created individual author profiles under Google Scholar Citations (GSC). This data was used by Webometrics Lab, Spain to create a ranked list of 4000+ institutions in 2016, based on citations from only the top 10 individual GSC profiles in each organization. (GSC excludes the top paper for reason,s explained in the text; the simple selection procedure makes the ranked list size-independent as claimed by the Cybermetrics Lab). Using this data (Transparent Ranking TR, 2016), we find the regional and country-wise distribution of GS-TR Citations. The size-independent ranked list is subdivided into deciles of 400 institutions each and the number of institutions and citations of each country obtained for each decile. We test for correlation between institutional ranks between GS-TR and the other ranking schemes for the top 20 institutions. Finally, we discuss our results in the context of questions like (1) Is it necessary to have one more global ranking scheme? (2) What are the likely benefits of the GS size-independent formulation? (3) What are the likely sources of error? and (4) Whether a truncated sample as in GS can indeed give a representative ranking acceptable at the global level?

**Keywords:** Google Scholar Citations, Citation, Open indicators, World Ranking of Universities.

## INTRODUCTION

The Transparent Ranking[1] by Cybermetrics Lab, beta version, was produced with the objective of trying to validate the use of Google Scholar Citations as a basis for obtaining a performance-based comparator list of research institutions and universities. It offers the possibility of open performance ranking and indicators without recourse to data that is behind paywalls, such as the Web of Science or Scopus data. Google Scholar Citations lists self-created profiles of individuals which display an individual's papers and their citations, machine updated by Google Scholar. Google Scholar does not state how many original journal sources it uses and additionally skims information off the net. As a result, GS may reflect non–peer-reviewed papers, both as source items as well as citing papers. This has two aspects. One relates to the 'quality', or lack thereof, of non-peer reviewed papers. For a long time now, the judgment of peers who are experts in the subject has been accepted as conferring acceptability to a scholarly work. This is augmented by having more than one peer review (two or three) to lend greater credence to the process of validation. Lately, however, more scientists are publishing on non-standard platforms, such as online journals and open archives such as ArXiv or SSRN. To restrict evaluation to only peer-reviewed journals would mean that all these additional papers would be missed. Finally, peer-reviewed journals are often subscription–based (alternatively Open Access) and therefore not accessible to a wide section of people. The use of the open format for Google Scholar based ranking of institutions implies that anyone can download the data and verify the calculations, whereas indicators which are not transparent and are based on priced databases are difficult to verify.

Google scholar has been widely written about and compared to other ranking schemes.[2-11] However, the new GS-TR size-independent Google Scholar Transparent Ranking has not been derived directly from Google Scholar, but from Google Scholar Citations (GSC).[12] With GSC Google rolled out a major enhancement in 2012, with the possibility for individual scholars to create personal "Scholar Citation profiles". These public author profiles are editable by the authors themselves.[13] Individuals, logging on through a Google account with a bona fide address usually linked to an academic institution, can now create their own page giving their fields of interest and citations. Google Scholar automatically calculates and displays the individual's total citation count, h-index and i10-index. According to Google, "three-quarters of Scholar Search results pages show links to the authors' public profiles" as of August 2014.[14,15] Prathap[16] has shown from the same data that the greater the scientific wealth of a nation, the more it will concentrate it in a few premier institutions.

In the following sections we will first describe the data and then examine how institutions of different countries are distributed in terms of citations as computed according to Google Scholar's Transparent Ranking (GS-TR) 'size-independent' methodology. Secondly, we will obtain the rank correlation of GS-TR ranks and those of other popular ranking schemes. Results are presented, followed by a section on the discussion.

### Data

Data was taken from a list of institutions ranked by Google Scholar Citations that appeared on the Webometrics website in July 2016.[1] The Transparent Ranking or Ranking Web of Universities was created by the Cybermetrics Lab (Spanish National Research Council), with the objective of testing the validity of using Google Scholar Citations (GSC) to rank universities and research institutions. It started in a small way in 2004 and has updated the rankings every six months, providing information about the performance of institutes all over the world. Currently, it is in an experimental state (beta).[16-21]

The methodology of the TR Ranking uses individual author profiles on Google Scholar Citations which have standardized (official) institution names and an official e-mail address. GSC automatically updates the citations in the individual profiles. There are at present about 1million profiles from 5000 institutions in Google Scholar Citations.[1] This data is nonetheless incomplete since people are required to voluntarily create their profiles on GSC, which implies that not every scientist/author has a profile. People are also required to make their profile public, for the profile to be included in the evaluation procedure. According to the Cybermetrics Lab, the data is large enough to give a representative ranking of world universities in spite of the incompleteness.

GSC-TR uses an unusual procedure in its evaluation. Not all the profiles in GSC are used. To make the ranking size-independent, only the top-ten (but excluding the first) profiles from an institution are selected and the citations aggregated over the remaining nine profiles to obtain the institutional citation. (The topmost profile is eliminated to exclude possible outliers). The citation aggregates are then used to rank the institutions. In other words, Google Scholar focuses on a sample of elite scientists to arrive at the ranking. It is similar in spirit to the Nature index,[22] which also focuses on an elite set of journals (and thereby elite scientists). At the same time, since it takes the same number of profiles from each institute, large or small, the resulting ranking is size-independent.

According to TR, steps are taken to deal with problems such as the presence of duplicate profiles, etc.[1] In the latest edition, the Ranking Web of Universities contains 4132 Institutions with an aggregate of 174,915,125 citations. This makes it one of the largest ranking exercises in the world. Harvard University has the highest rank with 1389765 citations. The country where the institution is located is also listed, making it possible for obtaining country-level indicators.

### Objective

Our objective in this paper is to gauge national citation performance based on the upper echelons within individual GSC profiles in an institution. Since the exercise by Cybermetrics Lab is in beta, our results will also be subject to the same caveats, i.e., uncertainties introduced by the incompleteness of the data and sampling technique described above. It is not clear for what period the citations have been collected, but possibly the period begins from the start of GSC to July 2016. The validity of the GS-TR list is tested by rank correlation with other currently accepted ranking schemes, using the top 20 institutions (Table 1).

## METHODOLOGY

We have divided the data on institutional ranking into percentiles (deciles of 400 institutions each; a few remaining institutions being grouped as 'Extra'). The top decile contains 400 best performing institutions with the highest citations. The country labels indicate how these top institutions are distributed among different countries. The 400 institutions in the second decile are a notch below those in the top decile and so on, progressively declining in quality till the 10$^{th}$ decile.

We have

**Table 1: Top 20 institutions from the Ranking Web and ranks from other schemes (2015-16).**

| Institutions | Country | Ranking Web July 2016 | THE 2015-16 | QS 2015-16 | ARWU 2016 | Leiden 2011-14 | SCIMAGO 2016 |
|---|---|---|---|---|---|---|---|
| Harvard University | USA | 1 | 6 | 2 | 1 | 1 | 3 |
| Stanford University | USA | 2 | 3 | 3 | 2 | 7 | 7 |
| Johns Hopkins University | USA | 3 | 11 | 16 | 101-150 | 5 | 13 |
| University of California Berkeley | USA | 4 | 13 | 26 | 3 | 16 | 26 |
| University of Chicago | USA | 5 | 10 | 10 | 10 | 82 | 106 |
| Massachusetts Institute of Technology | USA | 6 | 5 | 1 | 5 | * | 10 |
| University of Cambridge | UK | 7 | 4 | 3 | 4 | 18 | 19 |
| Michigan State University | USA | 8 | 99 | 164 | 101-150 | 89 | 181 |
| University of Oxford | UK | 9 | 2 | 6 | 7 | 13 | 14 |
| Columbia University New York | USA | 10 | 15 | 22 | 9 | 19 | 30 |
| University College London | UK | 11 | 14 | 7 | 17 | 15 | 23 |
| University of Michigan | USA | 12 | 21 | 30 | 23 | 3 | 11 |
| University of California San Diego | USA | 13 | 39 | 44 | 14 | 24 | 24 |
| Yale University | USA | 14 | 12 | 15 | 11 | 37 | 41 |
| McMaster University | Canada | 15 | 94 | 149 | 83 | 125 | 264 |
| Duke University | USA | 16 | 20 | 29 | 25 | 26 | 35 |
| University of California Los Angeles UCLA | USA | 17 | 16 | 27 | 12 | 12 | 16 |
| Temple University | USA | 18 | 351-400 | 601-650 | 301-400 | 367 | 395 |
| Princeton University | USA | 19 | 7 | 11 | 6 | 153 | 198 |
| Carnegie Mellon University | USA | 20 | 22 | 62 | 68 | 322 | 210 |

The basic characteristics of the ranking schemes mentioned here are shown in Table 2.

1) Examined the distribution of citations and institutions in countries in each of the deciles.
2) Tested for rank correlation between the institutions ranks assigned by GS-TR and other current ranking schemes.

(*The Google Scholar ranking is also referred to as the Ranking Web*)

The top 20 institutions (ranked by webometrics, Table1) were taken and their corresponding ranking from QS, LEIDEN, SCIMAGO, THE and ARWU were collected and tabulated (Table 1). As they differ in methodologies, the rankings can be fairly different in each case. The ranks are particularly different for the small universities since Web Ranking is size-independent. For example, the Webometrics ranks differ from all the others for Michigan State University and Temple University, both smaller universities. In some cases, e.g., Johns Hopkins University, all the ranks are fairly high except ARWU. In other cases, such as the Princeton University and Carnegie Mellon, THE, QS and ARWU are not very far from the GS rank, but Leiden and Scimago are widely divergent.

## RESULTS

We now characterize the properties of the GS-TR list based on Google Scholar Citations. There are a total of 4132 institutions and 174915125 citations. Harvard University has the highest rank with 1389765 citations, we note that among the top 20 institutions, 16 are from the USA, 3 from UK and 1 from Canada (we remind the reader that the citations are not the total citations accruing.

Citations accruing to an institution, but aggregated from only the top 10, but 1, Google Scholar profiles from that institution).

On dividing the list into deciles, we find that the number of countries within each decile ranges from about 30 to 80 (Figure 1), increasing approximately with decile number. 30 countries have at least one institution in the top decile (top 400 institutions). Countries with more institutions in the top deciles are more likely to have better performance.

### Citation Statistics

Citation statistics for different deciles are given in Table 3. Average citations per institute range from 272,923.6 citations in the top decile to 80.9 in the Extra decile (Table 3). The time period of the calculation of citations is not specified. The average and range of the citations in different deciles are seen in Figure 2.

| Table 2: Characteristics of selected schemes for ranking of world universities. | | | | | | |
|---|---|---|---|---|---|---|
| Ser No. | Ranking Scheme | Data used | Cost of Data | Partially Opinion based | Password controlled | Coverage |
| 1 | THE | WoS + Institutional data | Y | Yes | | Journal + Proceedings |
| 2 | QS | WoS + Institutional data | Y | Yes | | Journal + Proceedings |
| 3 | ARWU | WoS + Nobel award data | Y | No | | Journal + Proceedings |
| 4 | Leiden | WoS | Y | No | | Journal + Proceedings |
| 5 | Scimago | Scopus | Y | No | Yes | Journal + Proceedings |
| 6 | Ranking Web | Google Scholar Citations | Free | No | | Includes Web Citations |

WoS = Web of Science; THE = Times Higher Education Ranking, Britain, QS = Quacarelli-Symonds, Britain, Leiden Ranking, Netherlands, ARWU = Academic Ranking of World Universities, China, Scimago and Ranking Web, Spain.

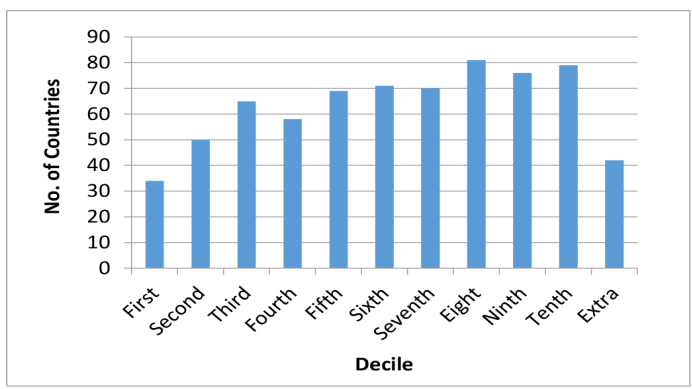

**Figure 1:** Number of countries in each decile of the ranked list of institutions.

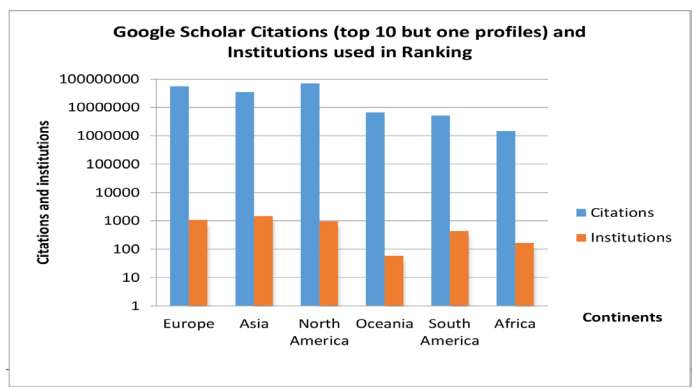

**Figure 3:** Distribution of Institutions and Citations in different World Regions.

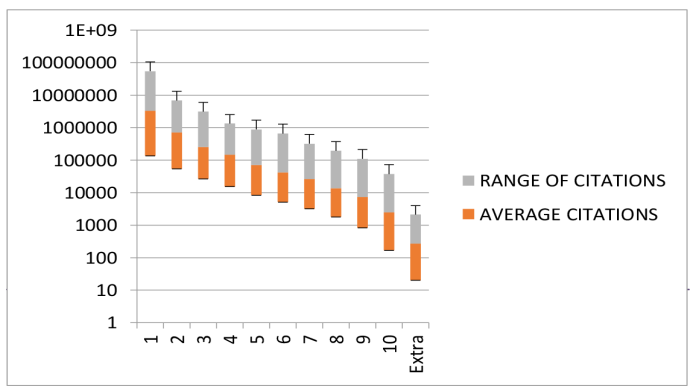

**Figure 2:** Box Plot showing average citation and range of citations in different deciles.

### Regional Distribution of GS-TR Citations and Institutions

Aggregating the citations and number of institutions for different countries into regions we obtain the regional distribution of citations (Figure 3). North America, Europe and Africa have more than 10,000,000 or 10 million GS citations each. Oceania, South America and Asia have between a million and 10 million GS citations each. The largest number of institutions is from Asia and the smallest from Oceania (<100).

### Distribution of a country's institutes in different deciles

Distribution of citations in different deciles is seen in Figure 4a and 4b. For every country, the number of institutions in the 1st, 2nd, 3rd, etc. deciles are seen in the column graph. A high performing country will have relatively more institutions in the top deciles. USA has largest number of institutions and citations in each decile.

In Figure 4a we see countries arranged in order of number of institutions (continued in Figure 4b). Each colour in a column stands for a decile (see legend). USA has overall 873 institutions in Google Scholar Citations (minimum citation 20) distributed in all the deciles. Countries with some institutions in the first decile get the highest contribution to citations from them. The average citation of an institute in the first decile is 272923.6 falling to 82674.5 in 2nd decile, 37094.4 in 3rd and so on (Table 3).

### USA has the largest share of institutions (21.13%) followed by India (7.94%)

Countries which have blank sections in the lowest part of a column in Figure 4a do not have any institutions in the first decile (e.g., Turkey, Iran, Poland, Malaysia, with no institutions in deciles 1 and 2, Columbia, Thailand and Argentina, Pakistan with no institutions in the first 3 deciles).

### Rank Correlation

From Table 4 the rank correlation appears high for THE, QS and ARWU. THE, QS also take into account perceptual variables such as reputation. ARWU includes awards, prizes in the rank computation. In contrast, Leiden and Scimago take scientometric indicators such as total papers and citations into account. These are found to be moderately correlated with GS, which is size-independent and based on top level citations.

**Table 3: Citation statistics for different deciles in ranked list of institutions, GS-TR.**

| Decile | No. of Institutions | Min Citation | Max Citation | Range | Avg. Citation/Inst | Std. Dev. | No. Countries |
|---|---|---|---|---|---|---|---|
| 1 | 400 | 125840 | 1389765 | 1263925 | 272923.6 | 165923.6 | 34 |
| 2 | 400 | 51950 | 125401 | 73451 | 82674.5 | 21168.6 | 50 |
| 3 | 400 | 25800 | 51947 | 26147 | 37094.4 | 7292.4 | 65 |
| 4 | 400 | 13802 | 25759 | 11957 | 19227.8 | 3395.9 | 58 |
| 5 | 400 | 8171 | 13785 | 5614 | 10755.8 | 1637.4 | 69 |
| 6 | 400 | 5060 | 8169 | 3109 | 6486.7 | 899.9 | 71 |
| 7 | 400 | 3121 | 5058 | 1937 | 4005.1 | 513.3 | 70 |
| 8 | 400 | 1731 | 3104 | 1373 | 2388.6 | 398.3 | 81 |
| 9 | 400 | 811 | 1728 | 917 | 1248.5 | 260.1 | 76 |
| 10 | 400 | 166 | 810 | 644 | 456.3 | 184.5 | 79 |
| Extra | 130 | 20 | 166 | 146 | 81.5 | 42.6 | 42 |

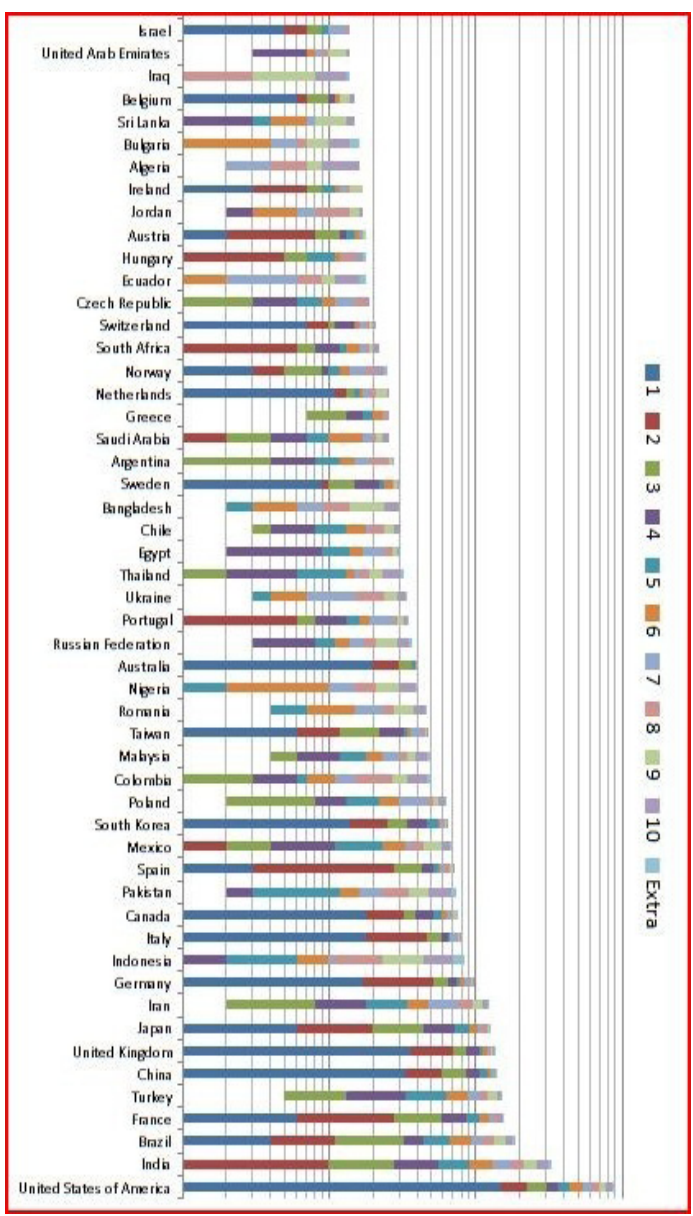
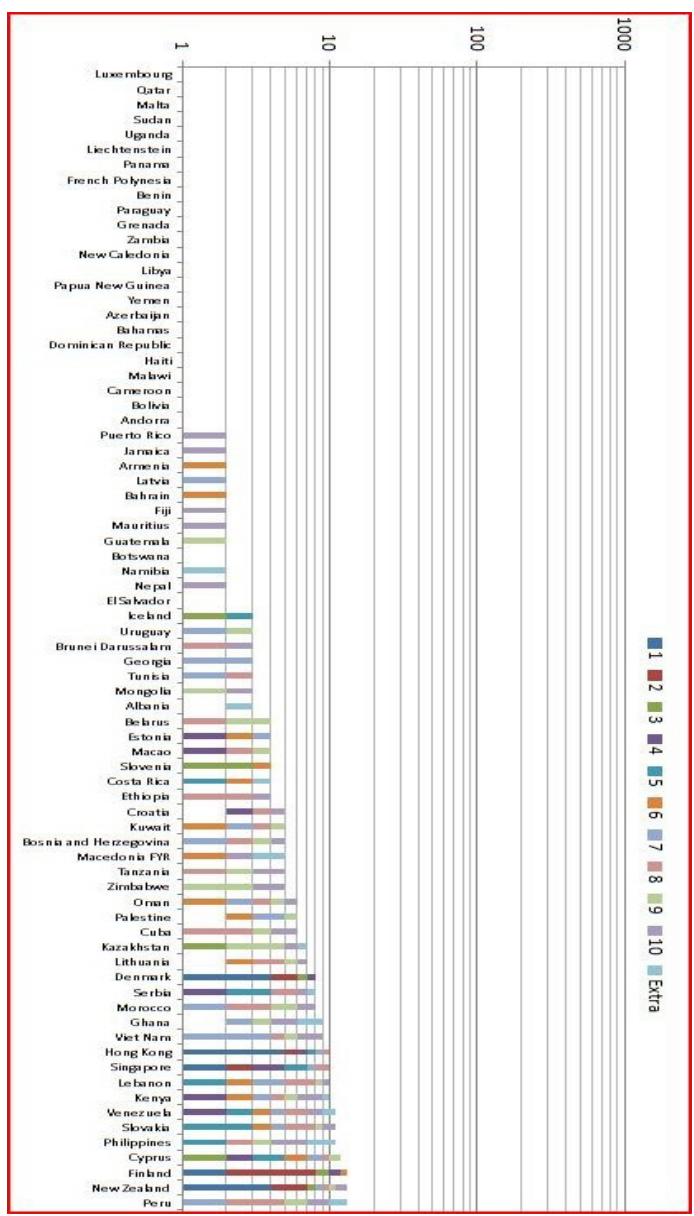

**Figure 4a:** Countries arranged in decreasing order of total institutes (right to left), colours indicating no. of institutes in each decile.

**Figure 4b:** Countries arranged in decreasing order of total institutes.

| Table 4: Rank Correlation of GS-TR and other ranking schemes, based on top 20 institutions. | | | | | | |
|---|---|---|---|---|---|---|
| **Pearson Rank Correlation** | **Ranking Web/ Google Scholar** | **THE** | **QS** | **ARWU** | **Leiden** | **SCIMAGO** |
| | Jul-16 | 2015-16 | 2015-16 | 2016 | 2011-14 | 2016 |
| **Ranking Web/ Google Scholar** Jul-2016 | 1 | | | | | |
| **THE** 2015-16 | 0.351355 | 1 | | | | |
| **QS** 2015-16 | 0.365987 | 0.997970 | 1 | | | |
| **ARWU** 2016 | 0.285652 | 0.935536 | 0.942091 | 1 | | |
| **Leiden** 2011-14 | 0.606561 | 0.710126 | 0.740549 | 0.717718 | 1 | |
| **SCIMAGO** 2016 | 0.554180 | 0.794356 | 0.810356 | 0.769585 | 0.908844 | 1 |

## DISCUSSION AND CONCLUSION

Earlier, performance-based global rankings of universities, or other institutions, whose members published academic or research papers, used diverse parameters including survey responses by experts, quality of teaching, internationalization in student and staff bodies, in addition to publication based performance evaluation in terms of number of papers and citations (e.g., Times Higher Education (THE) and Quacarelli-Symonds (QS). In these early exercises data was gathered from the universities to conduct the evaluation). ARWU, the Academic Ranking of World Universities from Jiao Tong University in Shanghai was the first ranking where data was taken from a non-partisan source, the Web of Science which recorded bibliographic details of journal articles and their citations. The ARWU gave weights to Nobel prize winners and Fields medal holders to rate the universities, in addition to the usual bibliometric indicators (papers, citations). The Leiden Ranking uses only bibliometric indicators (citation and collaboration) and ranks institutions in the first percentile (top 1% by citation) and decile (top 10%), It has both size-independent and size-dependent rankings. (A size-independent ranking can compare two institutions of different sizes). Other rankings, such as SCIMAGO from Spain, use SCOPUS which has a much higher coverage of journals than WoS for their data. Both Web of Science and Scopus have now introduced proceedings papers together with journal literature. The URAP from Turkey and the NTU from Taiwan are other recent entrants on the ranking scene. The Webometric Ranking differs from the others in several ways. It does not take its data[23-28] from a citation Index like WoS or SCOPUS, but directly from Google Scholar Citations on the web. Moreover, instead of cumulating over all individual citation profiles for an institute, it selects just nine author profiles and aggregates over these to obtain a representative citation for the institute.

The major advantage of using GSC is that it is freely available and accessible to anyone with an internet-enabled computer. Results obtained can be duplicated and verified. There is only a single indicator – citations – and there are no complicated calculations. The use of only the top profiles in an institution means that most of the information on citations is discarded and only the top edge retained to generate a rank order. How does the Google Scholar ranking compare with other schemas? Rank correlation shows that correlation is low with THE and QS. This may have been expected as they are partially survey-based and partially based on bibliometric indicators. The correlation with Scimago and the Leiden Ranking, which are based only on bibliometric indicators is better. In fact, correlation is best with the Leiden ranking (~ 0.7) which also has a size-independent variant.

We suggest that the procedure adopted by Webometrics in selecting their samples puts the focus on the 'cream' in each institution. This is also the segment with which academic 'reputation' is associated. However, unlike the Nobel Prize, it judges contemporary excellence and not historical reputation.

One of the criticisms faced by bibliometric evaluations of universities is that they emphasize research over teaching. Undoubtedly teaching is an important component of universities' responsibilities, but it is much harder to quantify. Good research, on the other hand, may also reflect good teaching as the two feed back into each other through students. The other criticism relates to the evaluation process. Peer review has been taken as the gold standard of evaluation, but in the case of world rankings, it would be clearly impractical for various reasons – the volume of information to be processed, the time required and the cost involved. Together with this, the fact that reviewers have expertise in small domains and cannot be expected to judge entire universities, especially when they are distributed all over the world.[22]

One can conclude that the Webometrics methodology has an advantage as it processes only a very small fraction of the full data. It is therefore economical in terms of data handling, time and cost. It is capturing something analogous to reputation or academic excellence by considering a few top-ranking members using bibliometric indicators rather than surveys. Finally, it is size-independent, a property that is very useful as sizes of universities are highly skewed. Otherwise, the ranking is dominated by the large established institutes and smaller institutes pass under the radar. Although Leiden and Scimago also use bibliometric indicators, papers and citations, they capture the overall citations of an institute and are seen to differ considerably from Webometric ranks in Table 3, even in the case of large and prestigious universities like Princeton and Carnegie Mellon. Finally, there is a movement toward open data in the academic world, which Google Scholar satisfies well.

Some of the disadvantages are the many errors that creep in while automatically collating the data on the net, the question of where to stop when taking non-peer reviewed literature into account and whether Webometrics will be able to capture the confidence of academics, managers and administrators the world over.

## ACKNOWLEDGEMENT

Aparna Basu thanks South Asian University for financial assistance as Guest Faculty.

## CONFLICT OF INTEREST

The authors submit that there were no financial or other considerations in the writing of this paper which may be construed as conflict of interest.

## ABBREVIATIONS

**GSC:** Google Scholar Citations; **TR:** Transparent Ranking; **WoS:** Web of Science.